# The polarization of two-level atom in the polychromatic weak field.

S.A.Pulkin, A.G.Antipov, A.Kalinichev, A.S.Sumarokov, S.V.Uvarova

Saint-Petersburg State University, Ulyanovskaya 1, Saint-Petersburg, Russia, 198504

## ABSTRACT

The numerical solution for polarization for two-level atom in polyharmonic field has been made. The analytical solution for partial case of symmetrical position of carrier frequency relative to transition frequency is possible. The results showed that the nonlinear features in polarization spectrum take place even for small amplitudes of comb – components for small frequency distance between them. It means that it is necessary to take into account nonlinear effects for interpretation of spectra in comb spectroscopy.

1. ## INTRODUCTION

The goal of the present work is to study the polarization spectrum of two-level system with homogeneously broadened line in a weak polyharmonic light field (comb). Comb spectroscopy is an expressive field of spectroscopy. The radiation spectrum consists of narrow equidistant peaks (comb spectrum). Nowadays comb spectroscopy is a fast developing area of spectroscopy allowing to detect with high sensitivity atomic and molecular lines in the wide spectral range with resolution limited by Doppler broadening for one–photon transitions. The main advantage of comb spectroscopy method is the possibility to detect simultaneously all spectral lines. The nonlinear effects are not taken into account in usual comb spectroscopy. We will show that the nonlinear features in polarization spectrum take place even for small amplitudes of comb – components for small frequency distance between them. The nonlinear effects arise when frequency distance between comb components is comparable (or less) with width of homogenously broadened line width. The Rabi – frequency of each comb component may be more less then line width. Recently we showed [1] that in counter-propagating combs the homogeneously broadened peaks arise on the wide Doppler counter.

## 2. ANALYTICAL SOLUTION FOR POLARIZATION OF TWO-LEVEL SYSTEM INTERACTING WITH POLYHARMONIC FIELD

Let's consider two-level atomic system driven by polyharmonic field with $(2K + 1)$ - monochromatic components:

$$E(t) = \frac{1}{2}\left(\left[E_{so} + \sum_{m=1}^{K} E_{sm}\left(e^{im\Delta_s t} + e^{-im\Delta_s t}\right)\right]e^{i\omega_{so} t} + c.c.\right),$$

where $\omega_{so} = (\omega_{2K+1} - \omega_1)/2$ is the middle (carrier) frequency of polyharmonic field, $\Delta_s = \omega_{j+1} - \omega_j$ – frequency distance between field component, $E_{sm}$ - the amplitude of m-component of field.

The density matrix formalism in the rotation wave approximation for non – moving atom was used. The system of differential equation for density matrix elements has the form:

$$\frac{d\rho_{12}}{dt} = -(\Gamma - i\delta)\rho_{12} + iV_{12}^* N_{12}$$

$$\frac{dN_{12}}{dt} = \lambda_{12} - \gamma N_{12} - 4Im(V_{12}\rho_{12}), \qquad (1)$$

where $\lambda_{12} = \lambda_1 - \lambda_2$ is the difference of pumping to the levels, $V_{12} = -d_{12}E(t)/\hbar$ is the matrix element of interaction energy in dipole approximation; $d_{12}$ - dipole transition momentum, $\delta = \omega_{21} - \omega_{s0}$ - the detuning of transition frequency from the middle frequency; $N_{12} = \rho_{11} - \rho_{22}$ - population difference; $\gamma = \gamma_1 = \gamma_2$, $\Gamma$ - the constants of longitude and transverse relaxations, the widths of level and line (fig.1).

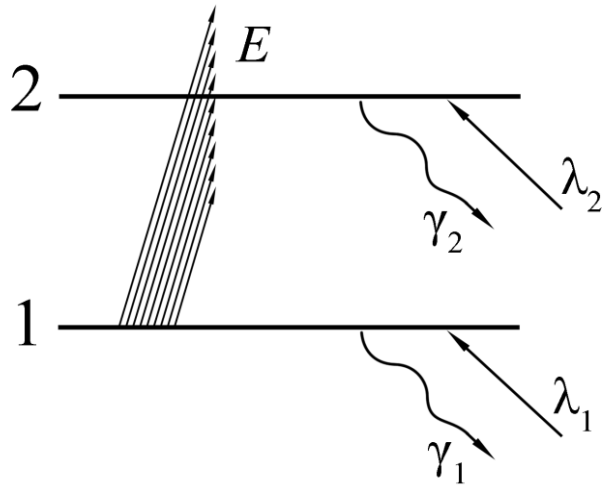

Figure1. The scheme of two-level atom interacting field E.

In the steady regime the non-diagonal elements of density matrix make periodical oscillations. The changing of the parameters of probe field, of the amplitudes of components $E_{sm}$ or frequency detuning $\Delta_s$ influence on the time dependences of population difference $N_{12}$ and on the non–diagonal elements of density matrix $\rho_{12}$. The period of oscillations $\rho_{12}(t)$ is determined by $\Delta_s$. For symmetrical case $\omega_{0s} = \omega_{21}$ the oscillations of $\rho_{12}$ are periodical and the susceptibility of atomic system is symmetrical relative of the middle frequency. For the non – zero detuning of field $\omega_{0s}$ relative the transition frequency $\omega_{21}$ the oscillations of imaginary and non - zero real part of $\rho_{12}$ have the same period. The population difference is the positive value, so there is inversionless system. The amplification on the some components takes place because re - pumping of energy from other components of pump field.

For symmetrical case the analytical solution is possible [2]. The time dependences for population difference and imaginary part of density matrix element have the form:

$$N_{12}(t) = \lambda_{12} \sum_{n_1=-\infty}^{+\infty} \cdots \sum_{n_K=-\infty}^{+\infty} \sum_{l_1=-\infty}^{+\infty} \cdots \sum_{l_K=-\infty}^{+\infty} \left( \frac{B_{n_1\ldots n_K l_1\ldots l_K}\, \gamma \cos f_{n_1\ldots n_K l_1\ldots l_K} \Delta_s t}{\gamma^2 + f^2_{n_1\ldots n_K}} - \frac{B_{n_1\ldots n_K l_1\ldots l_K}\, f_{n_1\ldots n_K} \sin f_{n_1\ldots n_K l_1\ldots l_K} \Delta_s t}{\gamma^2 + f^2_{n_1\ldots n_K}} \right),$$

$$\mathrm{Im}\,\rho_{12}(t) = \frac{\lambda_{12}}{2} \sum_{n_1=-\infty}^{+\infty} \cdots \sum_{n_K=-\infty}^{+\infty} \sum_{l_1=-\infty}^{+\infty} \cdots \sum_{l_K=-\infty}^{+\infty} \left( \frac{B_{n_1\ldots n_K l_1\ldots l_K}\, \gamma \sin f_{n_1\ldots n_K l_1\ldots l_K} \Delta_s t}{\gamma^2 + f^2_{n_1\ldots n_K}} + \frac{B_{n_1\ldots n_K l_1\ldots l_K}\, f_{n_1\ldots n_K} \cos f_{n_1\ldots n_K l_1\ldots l_K} \Delta_s t}{\gamma^2 + f^2_{n_1\ldots n_K}} \right),$$

where $B_{n_1\ldots n_K l_1\ldots l_K} = \prod_{m=1}^{K} J_{n_m}(-Z_{sm}) J_{l_m}(-Z_{sm})$, $Z_{sm} = 2\Omega_{sm}/m\Delta_s$, $\Omega_{sm} = -\frac{d_{21} E_{sm}}{\hbar}$, $f_{n_1\ldots n_K} = \Omega_{s0} - \sum_{m=1}^{K} m n_m \Delta_s$, $f_{n_1\ldots n_K l_1\ldots l_K} = \sum_{m=1}^{K} m(n_m - l_m)$.

The atomic polarization is determined by non-diagonal density matrix elements:
$$P(t) = d_{21} \rho_{12}(t) + \text{c.c.}$$

The polarization components oscillating on the frequencies $\omega_{sj} = \omega_{s0} \pm j\Delta_s$ are equal:

$$P(\omega_{sj}) = -\frac{d_{21}}{2} < iIm\rho_{12}e^{ij\Delta_s t} >_t,$$

where $< \cdots >_t$ is an averaging on time.

After averaging on time the real and the imaginary parts of polarization have the forms:

$$\mathrm{Re}(P_j) = \frac{\gamma \lambda_{12} d_{21}}{4} \sum_{n_1=-\infty}^{+\infty} \cdots \sum_{n_K=-\infty}^{+\infty} \sum_{l_1=-\infty}^{+\infty} \cdots \sum_{l_K=-\infty}^{+\infty} \left( \frac{B_{n_1..n_K l_1..l_K} \delta_{j,f_{n_1..n_K l_1..l_K}}}{\gamma^2 + f^2_{n_1..n_K}} - \frac{B_{n_1..n_K l_1..l_K} \delta_{j,-f_{n_1..n_K l_1..l_K}}}{\gamma^2 + f^2_{n_1..n_K}} \right),$$

(3)

$$\mathrm{Im}(P_j) = -\frac{\lambda_{12} d_{21}}{4} \sum_{n_1=-\infty}^{+\infty} \cdots \sum_{n_K=-\infty}^{+\infty} \sum_{l_1=-\infty}^{+\infty} \cdots \sum_{l_K=-\infty}^{+\infty} \left( \frac{B_{n_1..n_K l_1..l_K} f_{n_1..n_K} \delta_{j,f_{n_1..n_K l_1..l_K}}}{\gamma^2 + f^2_{n_1..n_K}} - \frac{B_{n_1..n_K l_1..l_K} f_{n_1..n_K} \delta_{j,-f_{n_1..n_K l_1..l_K}}}{\gamma^2 + f^2_{n_1..n_K}} \right),$$

where $\delta_{j,\pm f_{n_1..n_K l_1..l_K}}$ is the Kronecker symbol.

For weak fields and small intermodal distances when $\Delta_s$ is less or equal to $\Omega_{sm}$, the imaginary part of polarization has positive meanings for harmonics with number $-4<j<4$, where laser field exist. It means that these components are absorbed. For $|j|>4$ the imaginary part of polarization has negative value: the new components arise with amplitudes equal about 3/4 from positive parts with $-4<j<4$ (fig.2 a).

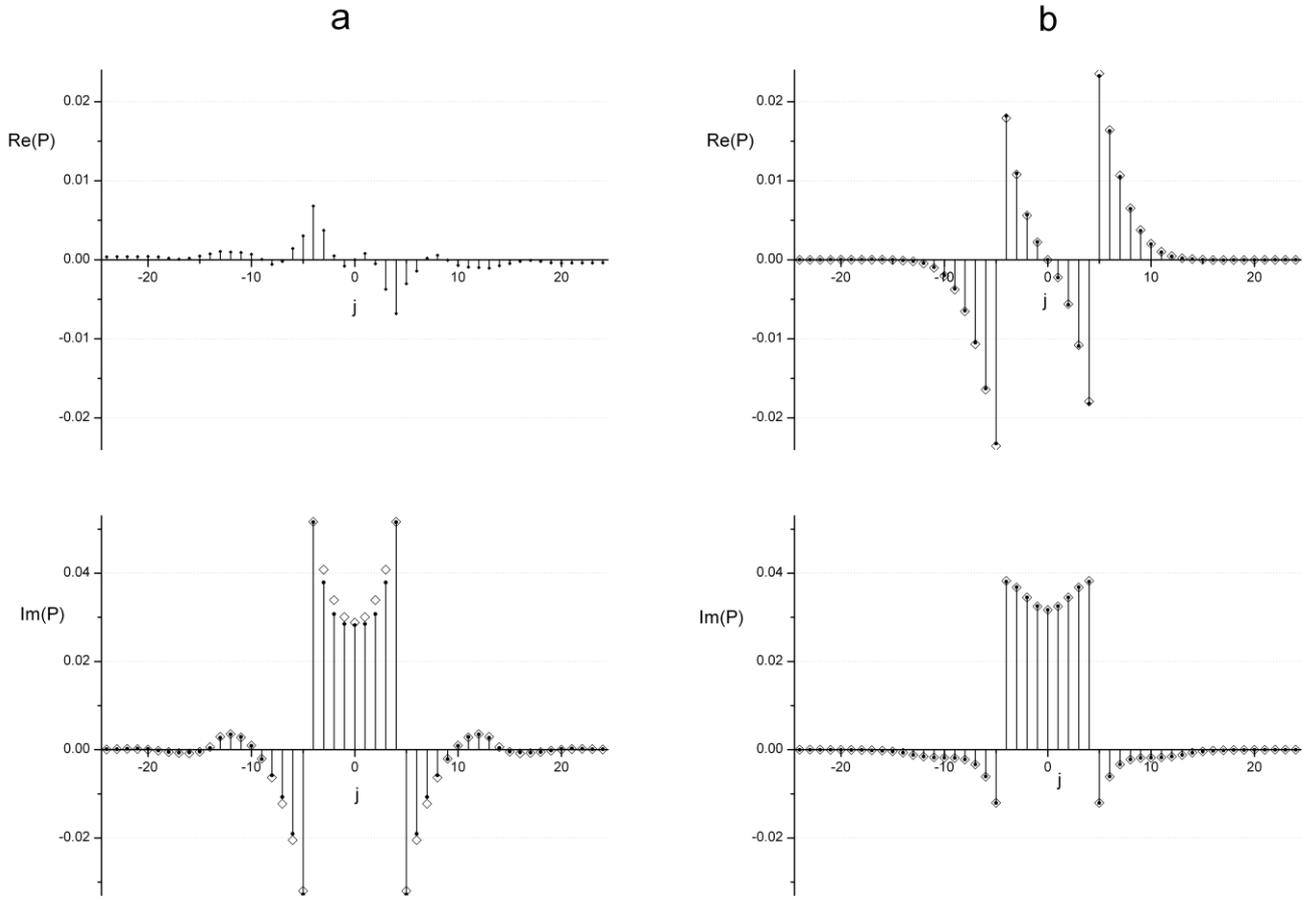

Figure 2 The real Re($P$) and the imaginary Im($P$) parts of polarization as functions of spectral number $j$. $K=4$, $\Omega_{sm}=0.2\Gamma$, $\Delta_s=0.01$ (a) and $\Delta_s=0.2$ (b), $\lambda_1=1$, $\lambda_2=0$, $\delta=0$. Optical pumping to lower level and inversion are absent and there is no detuning between central field frequency and the transition frequency (symmetrical field configuration). The diamond points show the results of direct formula (3) calculations while the small round points show the spectra of numerical solutions of the differential equation system (1).

For large inter-mode distances $\Delta_s > \Omega_{sm}$ the dependence of imaginary part of polarization has the usual Lorentz form as it shoved on Figure 3. The real part has the usual form too.

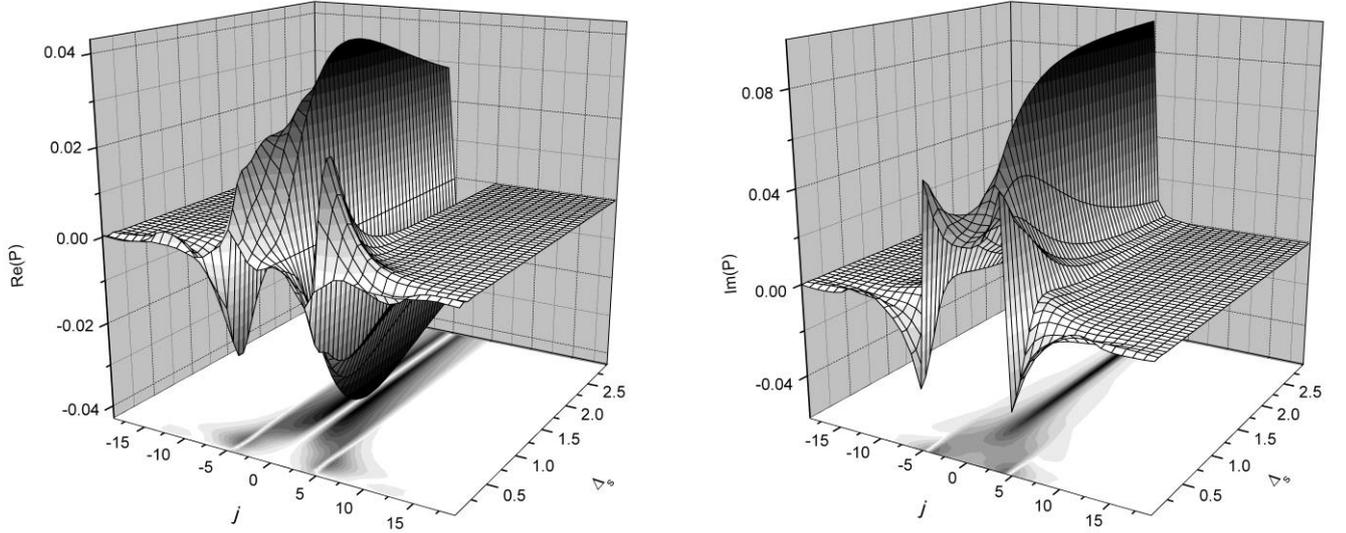

Figure 3. The dependence of real part of polarization Re($P$) and imaginary part Im($P$) on the number $j$ and inter-mode distance $\Delta_s$ for the following system parameters: equal amplitudes of 9-harmonic – field, $K = 4$, $\Omega_{sm} = 0.2$, $\lambda_1=1$, $\lambda_2=0$, $\delta=0$.

We checked analytical solution with numerical one and found good agreement [3, 4] (fig4).

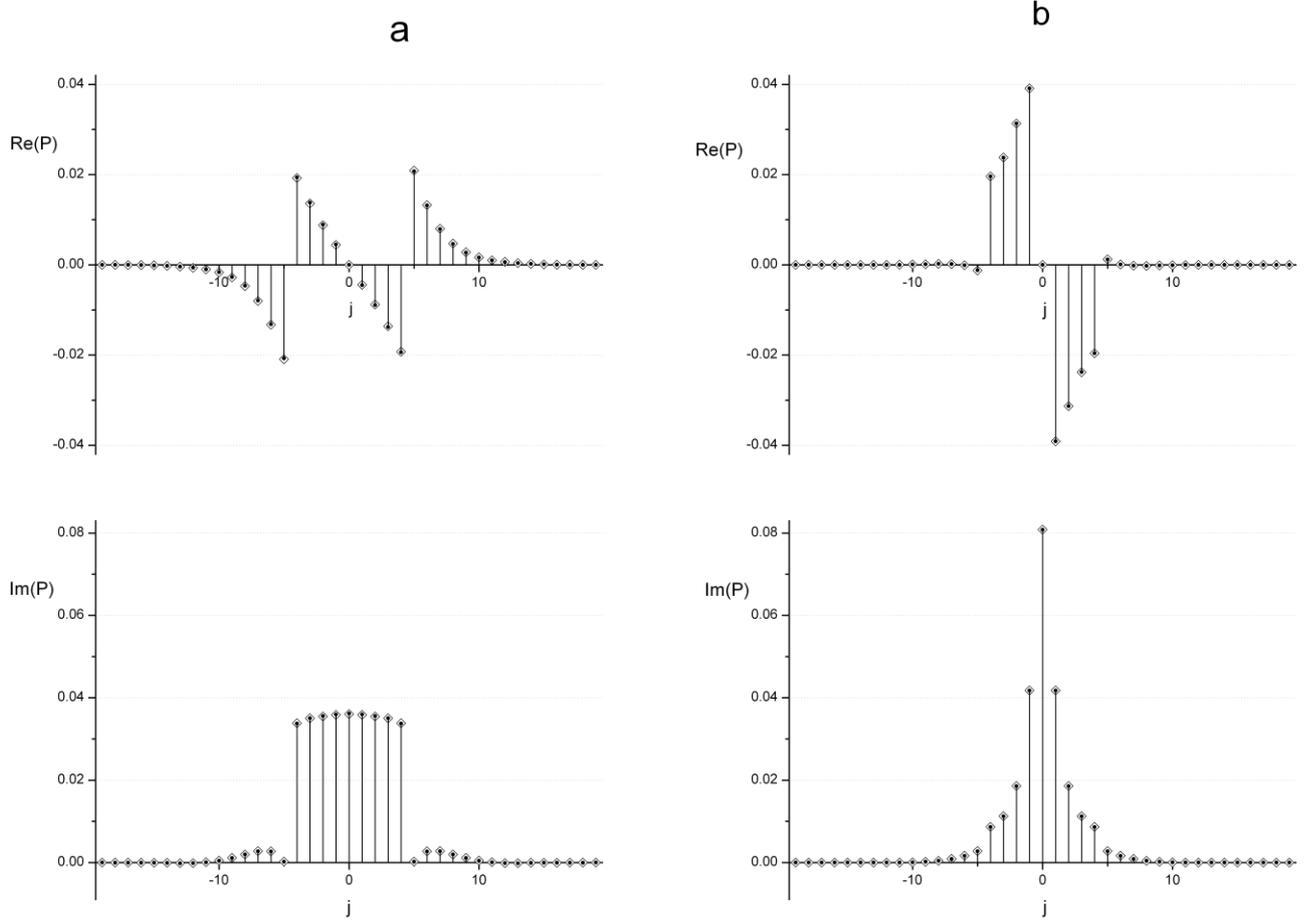

Figure 4. The real Re($P$) and the imaginary Im($P$) parts of polarization as functions of spectral number $j$. $K = 4$, $\Omega_{sm} = 0.2\Gamma$, $\lambda_1 = 1$, $\lambda_2 = 0$, $\delta = 0$, $\Delta_s = 0.3$ (a) and $\Delta_s = 1.0$ (b). The diamond points show the results of direct formula (3) calculations while the small round points show the spectra of numerical solutions of the differential equation system (1).

## CONCLUSIONS

The probability of nonlinear coherent processes in the case of polychromatic field is determined not only by the amplitude of driving laser field, but by the distance between comb-components. It connected with influence of components each another for small distances through nonlinear atomic medium. It is necessary to take into account the nonlinear phenomena for comb spectroscopy.